# Astro2020 Science White Paper

## Kiloparsec-scale Jets: Physics, Emission Mechanisms, and Challenges

**Thematic Areas:** ☐ Planetary Systems   ☐ Star and Planet Formation
☒Formation and Evolution of Compact Objects   ☐ Cosmology and Fundamental Physics
☐Stars and Stellar Evolution  ☐Resolved Stellar Populations and their Environments
☒Galaxy Evolution       ☒Multi-Messenger Astronomy and Astrophysics

**Principal Author:**
Name:           Eric S. Perlman
Institution:    Florida Institute of Technology
Email:          eperlman@fit.edu
Phone:          +1-321-674-7741

**Co-authors:** (names and institutions)

Eileen Meyer (UMBC), Jean Eilek (NMIMT), Sasha Tchekhovskoy (Northwestern), Kristina Nyland (NRL), Ivan Agudo (IAA, U. Granada), Stefi Baum (U. Manitoba), Martin Hardcastle (U. Hertsfordshire), Matthias Kadler (U. Wurzburg), Alvaro Labiano (CSIC-INTA), Herman Marshall (MIT/Kavli), Christopher O'Dea (U. Manitoba), Diana Worrall (U. Bristol)

**Abstract** (optional):

Jets are a ubiquitous part of the accretion process, created in AGN, by a coupling between the magnetic field near the central black hole and inflowing material. We point out what advances can be achieved by new technologies, concentrating on kiloparsec scales, beyond the Bondi radius, where accretion stops.  Here, jets profoundly influence their host galaxy and the surrounding clusters and groups, transporting prodigious amounts of matter and energies to scales of hundreds of kpc.  Basic questions still remain regarding jet physics, which new instruments can advance greatly.   The *ngVLA, LOFAR, JWST* and *LUVOIR*, as well as a *Chandra* successor, will give higher angular resolution and sensitivity.  This will allow us to probe the emission mechanisms and dynamics of jets, and search for links between these areas, magnetic fields, particle acceleration and high-energy emission mechanisms.  We stress the need for polarimetry in the X-ray and optical, critical to many of the most important questions in jet physics.  We hope to directly probe resolved, flaring components, which for the first time will allow us to reveal how jets respond to stimuli and link statics and dynamics.

Jets are a common feature of accreting systems, likely created in active galactic nuclei (AGN) near the central black hole by a coupling between the magnetic field and accreting material [59]. AGN jets transport energy and mass from sub-parsec central regions to Mpc-scale lobes, with a kinetic power comparable to that of the host galaxy and the AGN [55, 14, 4]. This White Paper concentrates on kiloparsec-scale jets, between the Bondi radius (where accretion stops) and the larger flow that can reach hundreds of kpc. Here, jets profoundly influence their host galaxy and the surrounding clusters and groups. Over the last decade, the synergy between the *VLA*, *HST*, *Chandra*, and *ALMA*, has allowed us to understand much better the kinematics and dynamics of kiloparsec-scale jets, but just as many new questions remain. These include:

1. Over what scales do jets maintain relativistic velocities, and where do they decelerate?
2. What maintains relativistic particles and magnetic fields in the jet on kiloparsec scales?
3. What governs jet dynamics and structure, and interactions with surrounding media?
4. What is the emission mechanism of the jet at optical, X-ray and higher frequencies?
5. How are particles accelerated locally, once the jet has left the host galaxy's nucleus?

## I. The Need for New Observational Tools

*HST* and *Chandra* represented a drastic increase in angular resolution and sensitivity, resulting in the discovery of X-ray and optical jets. An equally large discovery space is available by further increasing the resolution and sensitivity, with full polarimetric capability. Ground-based 30m telescopes with adaptive optics will enable near-milliarcsecond resolution in the optical, as will *LUVOIR* in the UV. In radio, the *ngVLA* will feature high sensitivity, high angular resolution, and broad frequency coverage, capabilities *LOFAR* will bring those to lower frequencies. In X-rays, *Lynx* or another *Chandra* successor mission could feature improved resolution and sensitivity.

Much remains unknown about jets because they are faint and lack obvious diagnostics. For example, their matter content is poorly constrained because jets are completely ionized and emit continuum radiation only, so that spectral lines or other features [15, 40, 41] cannot be used. The emission mechanism of quasar jets in X-rays is also poorly constrained (Section III), as are dynamical processes. Polarimetry of quasar jets can help with both, but it is currently impossible in the X-rays, and difficult in the optical with *HST* (Section IV): the four brightest required 9-21 orbits each. Moreover, a 200-500 ks *Chandra* observation can obtain over 100 counts per knot for only a handful of jets (the threshold for spectral modeling), and the knots are usually unresolved in the X-rays, but resolved in UV/optical or radio – because *Chandra*'s angular resolution is an order of magnitude worse than that of *HST*, J*VLA* or *ALMA*. Making major progress on the dynamics and emitting processes of X-ray emitting regions, and even the thornier question of the jet's matter content [26] requires angular resolution <0.1" in X-rays.

With increased angular resolution and sensitivity, we could for the first time connect fine-scale physics, particle acceleration, dynamics and variability in resolved, flaring knots. To date, this has been done only in knot HST-1 in the M87 jet (Figure 1), the nearest one seen in the optical – in all other such knots the emission is confused with other regions. The shock model of HST-1 [46] connected these issues and agreed with proper motion data. Generalizing such models requires more examples. Recent work [32, 20, 5, 10] shows that several more distant jets may contain varying, kpc-scale regions that we cannot resolve currently. Higher resolution, higher sensitivity telescopes in the X-ray and optical/UV would allow this critical work to be done.

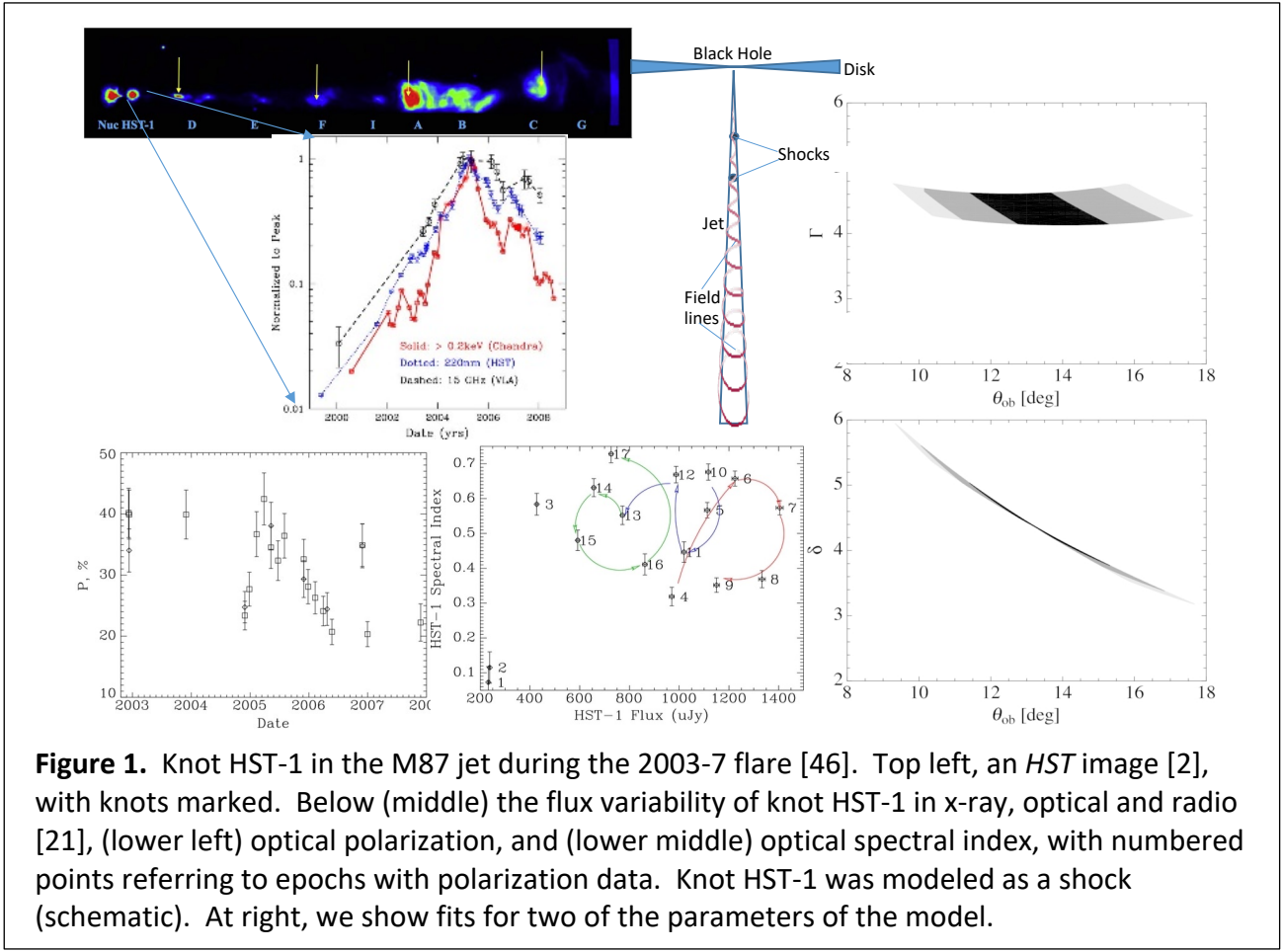

**Figure 1.** Knot HST-1 in the M87 jet during the 2003-7 flare [46]. Top left, an *HST* image [2], with knots marked. Below (middle) the flux variability of knot HST-1 in x-ray, optical and radio [21], (lower left) optical polarization, and (lower middle) optical spectral index, with numbered points referring to epochs with polarization data. Knot HST-1 was modeled as a shock (schematic). At right, we show fits for two of the parameters of the model.

**II. Radio, Optical and X-ray Imaging and Kinematics**

High-resolution, multi-wavelength maps have enabled a clear view of complex structures in nearby jets. As an example, Figure 2 shows deep imaging of the 3C 111 jet [9]. Notably, optical and X-ray emission maxima are often upstream from the radio peaks. This trend, also seen elsewhere [17, 53, 23], suggests different particle populations – as does the fact that the X-ray (and in some cases, optical/UV) are a separate spectral component even when spatially coincident.

While VLBI studies of jet kinematics on pc scales have been done for decades (e.g., [24, 28, 29]), studies on >100 pc scales with the *JVLA* [3], *HST* [41, 36, 37, 40] and *Chandra* have only been possible in a few cases due to the need for decades-long time baselines. Only 4 nearby, FR I jets have pc- *and* kpc-scale kinematic observations. All show speeds $\lesssim c$ on scales up to ~100 pc (projected) from the core. At larger scales, Cen A and M84 exhibit slightly faster, but still sub-*c* speeds, while M87 and 3C 264 show a stationary knot beyond which superluminal speeds and gradual deceleration with distance are seen. But in the powerful quasar jet of 3C 273, which has pc-scale speeds up to 15 *c* [28, 29], *HST* observations give an upper limit <2*c*[1] for knots >100 kpc from the core, ruling out the IC/CMB interpretation of the X-ray emission from these knots [35].

---

[1] For a jet with speed $\beta = v/c \geq 0.9$, inclination angles $\theta \lesssim 40°$ will lead to apparent superluminal speeds $\beta_{app} > 2$, as $\beta_{app} = (\beta \sin\theta)/(1 - \beta \cos\theta)$, where the maximum $\beta_{app} = \Gamma$ is observed at an angle $\theta = 1/\Gamma$.

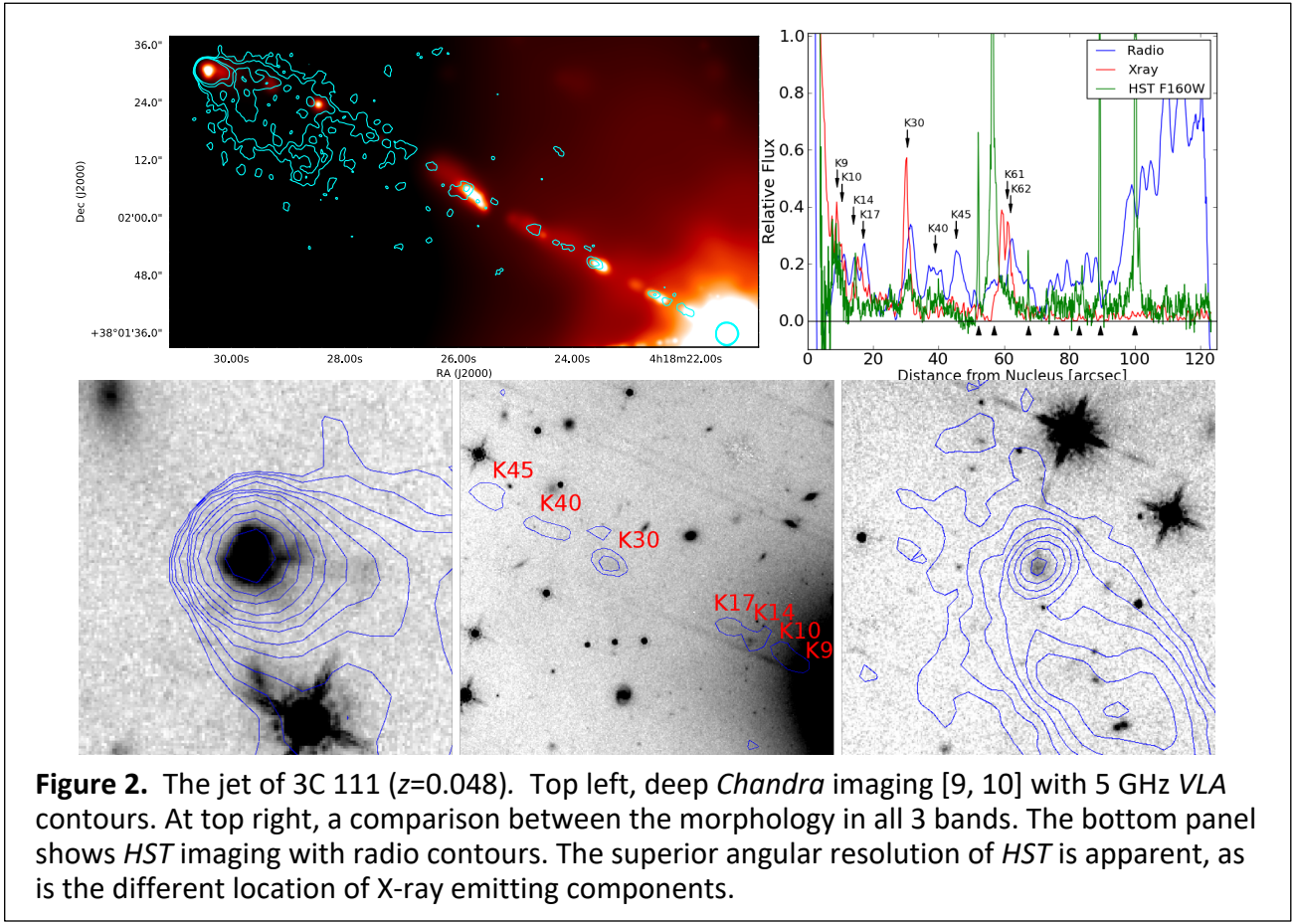

**Figure 2.** The jet of 3C 111 ($z$=0.048). Top left, deep *Chandra* imaging [9, 10] with 5 GHz *VLA* contours. At top right, a comparison between the morphology in all 3 bands. The bottom panel shows *HST* imaging with radio contours. The superior angular resolution of *HST* is apparent, as is the different location of X-ray emitting components.

We are only scratching the surface of what is possible with kpc-scale kinematic studies. Higher angular resolution and longer time baselines will fantastically advance this area, as long as very accurate geometric distortion corrections [60] are available for future optical and X-ray instruments' detectors. In the radio/mm, with *ALMA* and particularly *ngVLA* we will also be able to make spectacular progress. Note that work on M87 [42, 3] reveals that it is critical to obtain information in all bands, as the knots do not necessarily show the same speeds in each band.

## III. Emission Mechanisms in the UV, X-ray and Beyond

The first images from *Chandra* after its launch in 1999 showed X-ray emission from a kpc-scale jet in the quasar PKS 0637-752 (Fig. 2, [8]). This emission, over an order of magnitude brighter than an extrapolation of the radio continuum, was interpreted as inverse-Comptonization of CMB photons (IC/CMB; [58, 7]). This became the dominant paradigm for the next decade as X-ray emission from over 100 jets was discovered (e.g., [52, 30, 31, 33, 22]). These exciting images have revealed that X-ray jets are dominated by knot and hotspot emission. By contrast, in low-power, FR I jets, the optical/near-IR and X-ray fluxes usually fit on an extrapolation of the radio spectra (e.g., [47, 50, 51, 12, 13, 19]), consistent with a single-particle population synchrotron origin for the emission from radio through X-ray energies, though there are exceptions [39].

In the last decade, a controversy has emerged over the nature of the X-ray emission of quasar jets, with critical implications. While IC/CMB is a mandatory process, it requires bulk relativistic motion ($\Gamma\sim10–20$) and small viewing angles on kpc scales, plus an electron energy distribution extending down to $\gamma<30$, much lower than the $\gamma\sim100–1000$ traced by GHz emission. In view of the low radiative efficiency of such electrons, the X-ray flux requires high power, sometimes exceeding the black hole's Eddington luminosity [1, 18, 11, 34, 61, 62]. The detection of high polarizations in the optical – often part of the same spectral component ([6,9]; § III) – as well as the lack of the required GeV emission from knots [35, 36, 37, 5], necessitates an alternative interpretation, except perhaps at very high redshift where the CMB is enhanced [53].

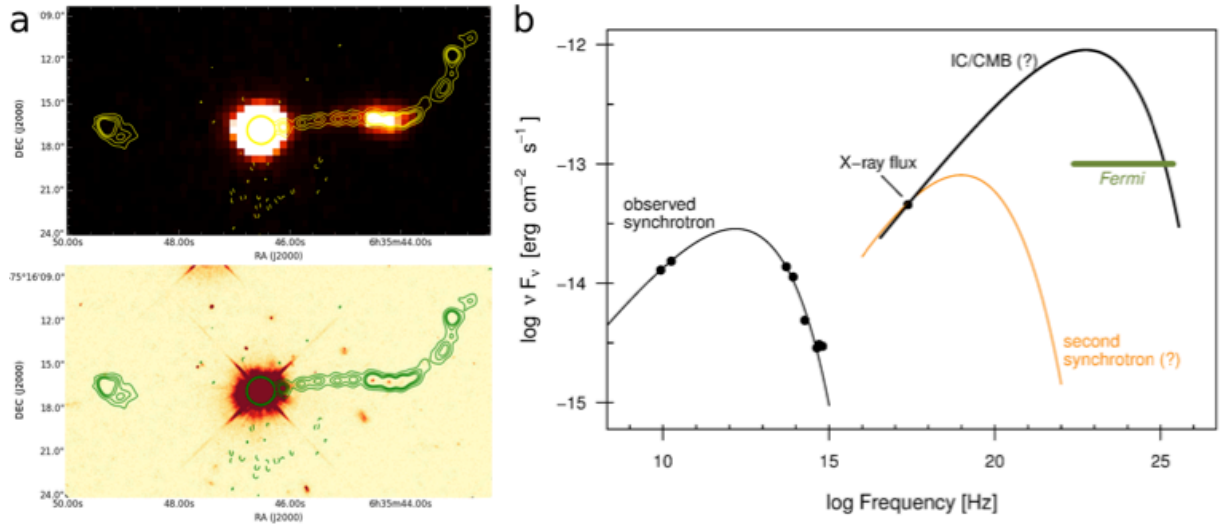

**Figure 3.** (a) X-ray and optical images of the 100-kpc long jet of the quasar PKS 0637-752, with 17 GHz radio contours overlaid. The unexpected X-ray detection of the jet led to the original suggestion that quasar jets remain highly relativistic to kpc scales. (b) At right, the total radio-X-ray SED (black points). The X-ray emissions require a separate spectral component, but the *Fermi* upper limits rule out an IC/CMB interpretation. [36, 34].

Alternatively, quasar jets' X-ray emissions may be synchrotron emission from high energy electrons, lying in a separate, high-energy population [22, 18]. This does not require a highly relativistic kpc-scale jet, viewed at small angles, nor does it make heavy demands on the kinetic power [1]. It also naturally explains the high polarizations in the optical [6], X-ray/radio flux morphology differences [8] and lack of detected GeV gamma-ray emission, and is also attractive in less-beamed objects such as 3C 111 [9], where the jet's synchrotron spectrum appears to extend up to at least 10 keV. But synchrotron X-ray emissions imply efficient, *in situ* particle acceleration reaching at least $\gamma\sim10^8$ hundreds of kpc from the central engine, seemingly far away from dynamical interactions that could trigger shocks or other accelerating events. This is validated by the finding of variable X-ray emissions in several kpc-scale knots [46, 32, 20, 5, 10], however, and in addition it is likely that future, higher-sensitivity observations of faint X-ray knots will reveal additional, low-level variability (see *e.g.,* [16] for the state of the art in Cen A).

### IV. Radio and Optical Polarimetry – Jet Structure and Particle Acceleration

In low-power FR Is, polarimetry has addressed fundamental issues, including the 3-D magnetic field configuration, and the role of dynamics and fields in particle acceleration. In the M87 jet, differences between polarization and morphology [47], filamentary structure and helical undulations [2] and flatter optical spectra in the knots [48], all suggest the acceleration of high-energy particles via turbulence, reconnection or stochastic processes in shocks. Modeling of the X-ray emission indicates that particle acceleration occurs in only a small fraction of the jet that decreases with distance from the nucleus [51]. Other FR Is also exhibit energetic stratification: some show optical polarization changes near X-ray knots [12, 13], but less strong differences exist in others [49, 50]. In two, we modeled in detail the evolution of the Stokes parameters in shocks [13, 49], and in one we saw a collision between superluminal shock components [38].

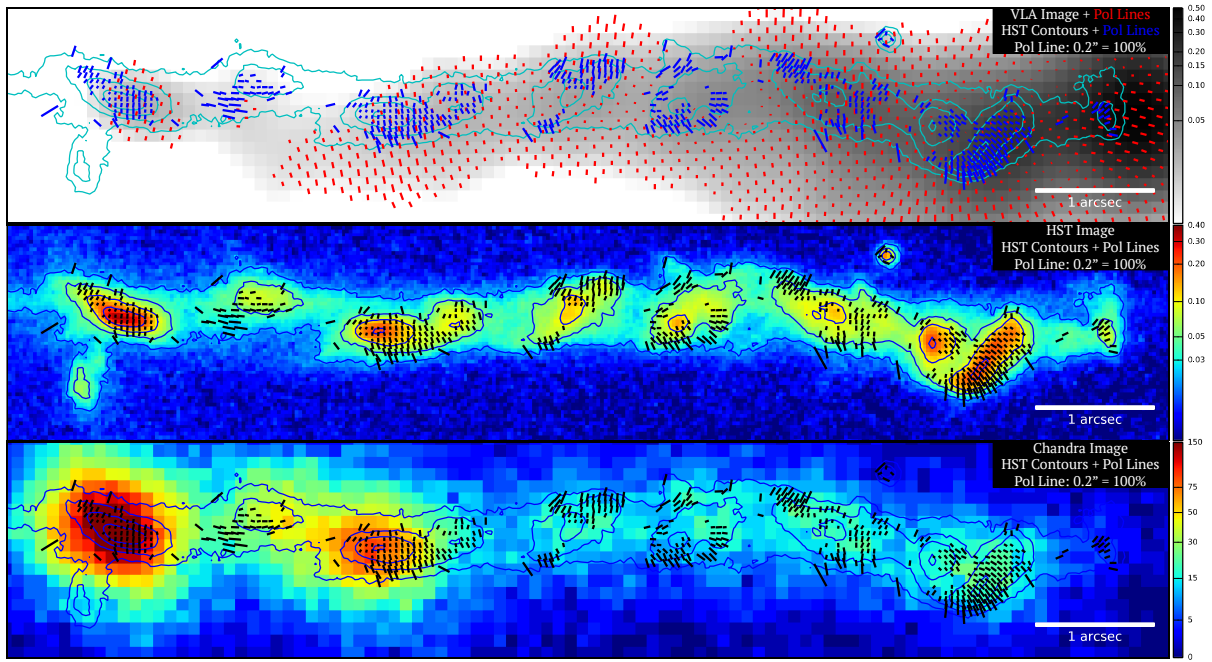

**Figure 4.** Optical and radio polarimetry and X-ray imaging [10] of the 3C 273 jet. The nucleus is not shown. Optical and radio polarimetry show vastly different results, suggesting that particle acceleration only occurs in a small part of the jet volume. Also seen is an anti-correlation between optical and X-ray flux and polarization and a long, low-polarization channel.

Recent *HST* polarimetry of four quasar jets offers the same promise as in FR Is, as well as an extra possibility of diagnosing the high-energy emission mechanism, as IC/CMB should be unpolarized [61, 27]. Our work [6, 10] shows highly (up to 40%) polarized optical knots, and in three of four cases the optical and X-ray emissions are on the same spectral component. This strongly disfavors the IC/CMB mechanism and favors a second, high-energy synchrotron emitting electron population. Interestingly, though, that population seems to be both within knots and well distributed throughout the jet (see e.g., Figure 4). This suggests diverse particle acceleration processes despite the location, tens to hundreds of kpc away from the host galaxy.

**Conclusion.** Optical and X-ray imaging polarimetry, combined with variability of resolved knots, offer the best way to test models for quasar jets' X-ray emission, where synchrotron model is favored for most, and dynamics. Future work can link these processes to magnetic fields and jet dynamics. The future is bright, but full understanding requires future instruments to have high sensitivity and angular resolution in *all* wavebands, from radio through hard X-rays.